\documentclass[twoside,twocolumn]{aastex63}

\usepackage{natbib}
\usepackage{graphicx}
\graphicspath{{./Figures/}}

\usepackage{comment}
\usepackage{printlen}
\usepackage{amsmath}
\usepackage{xcolor}

\usepackage{enumitem}
\setlist{noitemsep}

\usepackage{soul}

\usepackage{hyperref}
\begin{document}

\title{
    Spatial and temporal analysis of
    3-minute oscillations in the chromosphere
    associated with the
    X2.2 Solar Flare on 2011 February 15
}

\author{Laurel Farris}
\affiliation{New Mexico State University}

\author{R. T. James McAteer}
\affiliation{New Mexico State University}

\begin{abstract}
    3-minute oscillations in the chromosphere
    are attributed to both slow magnetoacoustic waves
    propagating from the photosphere, and to
    oscillations generated within the chromosphere itself
    at its natural frequency as a response to a disturbance.
    Here we present an investigation of the spatial and temporal
    behavior of the chromospheric 3-minute oscillations
    before, during, and after the SOL2011-02-15T01:56 X2.2 flare.
    Observations in ultraviolet emission centered on
    1600 and 1700 Angstroms obtained at 24-second cadence
    from the Atmospheric Imaging Assembly
    on board the Solar Dynamics Observatory
    are used to create power maps as functions of both space and time.
    We observe higher 3-minute power during the flare,
    spatially concentrated in small areas
    $\sim$10 pixels ($\sim4\arcsec$) across.
    This implies that the chromospheric plasma is not oscillating
    globally as a single body.
    The locations of increased 3-minute power are consistent with
    observations of HXR flare emission from previous studies,
    suggesting that these small areas are manifestations of the
    chromosphere responding to injection of energy by nonthermal particles.
    This supports the theory that the chromosphere oscillates
    at the acoustic cutoff frequency in response to a disturbance.
\end{abstract}


\section{Introduction}\label{introduction}

Most of the radiative energy associated with solar flares
is \{emitted\} from the chromosphere
in the form of optical and UV emission, but the mechanism of energy
transport from the magnetic reconnection site to the chromosphere
and subsequent conversion to other forms remains unclear
\citep{Hudson2007, Hudson2009}.
The flaring chromosphere has been observed
to oscillate seemingly in response to an injection of energy,
suggesting that such oscillations may reveal
something about the nature of energy deposition and conversion
associated with flares
\citep{Kumar2006, Monsue2016, Milligan2017}.

3-minute oscillations in the chromosphere
in intensity have been observed in the quiet sun internetwork
\citep{Orrall1966} and over active regions \citep{Beckers1969, Wittmann1969},
and have been attributed to the upward propagation of slow
magnetoacoustic waves that originate in deeper layers of the solar atmosphere
\citep{OShea2002, Tian2014}.
At the base of the chromosphere, the acoustic cutoff frequency
$\nu_{0} \approx 5.6$ mHz, which corresponds to a period of $\sim$3 minutes,
effectively creates a barrier to waves with propagation
frequency $\nu < \nu_{0}$.

It has been theoretically predicted that the chromospheric plasma will
naturally oscillate at the acoustic cutoff frequency in response to a
disturbance, regardless of the frequency of the disturbance itself and that
this applies to both continuous disturbance (such as propagating waves from
the photosphere) and impulsive disturbances, such as a flare
\citep{Fleck1991, Sutmann1995a, Sutmann1995b, Sutmann1998, Chae2015}.

\cite{Kumar2006} studied Dopplergrams during two flares and observed an
enhancement in velocity oscillations at frequencies between 5 and 6.5 mHz. The
observed
enhancements occur in close proximity to the HXR
source as derived from RHESSI images, indicating the enhancement was a
response to energy injection by non-thermal particles.
\cite{Kosovichev2007} analyzed \ion{Ca}{2} H emission and found high-frequency
oscillations
exceeding the acoustic cutoff frequency
associated with a
flare at chromospheric heights over a sunspot umbra and inner penumbra.
Propagation speeds $\approx$ 50--100 km s$^{-1}$ exceeding the local sound
speed, along with spatial localization in areas of high magnetic field strength
supported the interpretation of the oscillations as signatures of traveling MHD
waves.
\cite{Brosius2015} studied UV sit-and-stare spectra of an M-class flare in
\ion{Si}{4}, \ion{C}{1}, and \ion{O}{4} lines, and reported four complete
intensity fluctuations with periods around 171 seconds, which was attributed to
energy injection in the chromosphere by non-thermal particle beams. Using
high-resolution spectra from the
\textit{Interface Region Imaging Spectrograph}
(\textit{IRIS}; \cite{DePontieu2014}),
\cite{Kwak2016} found a chromospheric response to a downflow event at a
3-minute period, following the predictions by \cite{Chae2015}.

Small-scale fluctuations known as quasi-periodic pulsations (QPPs)
are observed in flare emission for wavelengths ranging from radio through
hard X-rays, with periods ranging from milliseconds to tens of minutes,
and persist throughout the duration of flare development and evolution
(for recent reviews, see \cite{Nakariakov2009} and \cite{VanDoorsselaere2016}).
The mechanism by which QPPs are generated
is thought to be either periodic episodes of magnetic reconnection or a
signature of MHD waves in the ambient plasma.
Particularly when observed in
thermal emission from the chromosphere, QPPs may provide valuable diagnostics
for the energy transportation and injection processes \citep{Inglis2015}.
There are relatively few studies of QPPs in thermal emission due to the low
modulation depth of the signal at these wavelengths \citep{Hayes2016}.

\cite{Monsue2016} analyzed data from observations of two flares (M- and
X-class) by Global Oscillation Network Group (GONG) in H$\alpha$ emission.
Analysis of emission integrated over the active regions revealed an enhancement
in power for a range of frequencies between 1.0 and 8.3 mHz (periods between
$\sim$16.7 and $\sim$2.0 minutes, respectively).
The same analysis on a
subregion containing only the inner flare region revealed a suppression in
power for all frequencies during main phase of flare, but enhancement at lower
frequencies between 1 and 2 mHz immediately before and after the flare. The
spatial and temporal information helps identify when and where energy
transformation to other forms occurs, and the enhanced low-frequency power
prior to a flare may be a signature of an instability in the plasma.

\cite{Milligan2017} analyzed Lyman-$\alpha$ emission from \textit{GOES},
full disk Lyman continuum data from \textit{SDO}/EVE,
and thermal UV emission from \textit{SDO}/AIA integrated over the AR
during the X2.2-class flare that occurred on 2011 February 15 in NOAA AR 11158.
The largest power enhancement occurred during the flare for periods between 100
and 200 seconds, similar to rate of energy injection observed in HXR emission
from RHESSI associated with the flare.
There was also an increase in power around the 3-minute period, which was not
reflected in the HXR flare data from RHESSI, and was thought to be independent
of the energy injection rate and intrinsic to the plasma itself.
This was attributed to the chromosphere naturally responding at the acoustic
cutoff frequency, though the spatial size and location of this response within
the active region remains unknown.
Here we
expand on the work of \cite{Milligan2017}
and present the spatial and temporal evolution of 3-minute power in the
chromosphere during the 2011 February 15 X-class flare, using data from the
Atmospheric Imaging Assembly (AIA; \cite{Lemen2012}) on board the \textit{Solar
Dynamics Observatory} (SDO; \cite{Pesnell2012}). These data allow the
computation of spatially resolved power maps centered on the frequency of
interest. Observations are described in \S\ref{data}, followed by analysis
methods in \S\ref{analysis}. Results are presented and discussed in
\S\ref{results}. We conclude in \S\ref{conclusions} with key preliminary
findings and plans for the continuation and development of this work.


\section{Observations and data reduction}\label{data}

The 2011 February 15 X2.2 flare occurred in NOAA active region (AR) 11158
close to disk center
during solar cycle 24 (SOL2011-02-15T01:56). The AR was composed
of a quadrupole: two sunspot pairs (four sunspots total). The X-flare occurred
in a delta-spot composed of the leading spot of the southern pair and the
trailing spot of the northern pair \citep{Schrijver2011}.
It started at 01:44UT, peaked at 01:56UT, and ended at 02:06UT,
as determined by the soft X-ray flux from the
\textit{Geostationary Operational Environmental Satellite}
(\textit{GOES}-15; \cite{Viereck2007}).

\textit{SDO}/AIA obtains full disk images throughout the solar atmosphere,
including two filters that provide measurements of thermal UV emission from the
chromosphere. The 1700\AA{} channel primarily contains continuum emission from
the temperature minimum, and the 1600\AA{} channel
is sensitive to
both continuum emission in the upper photosphere
and the \ion{C}{4} spectral line in the transition region.
Both UV channels have a cadence of 24 seconds and
spatial size scale of {0.6\arcsec} per pixel.

Data from the Helioseismic and Magnetic Imager (HMI; \cite{Scherrer2012}), also
on board \textit{SDO}, is used to study possible correlations
between magnetic field strength and oscillatory behavior in the chromosphere.
HMI obtains full disk data in four types of filtergrams: line of sight
magnetograms, vector magnetograms, Doppler velocity, and continuum intensity.
Each data product is centered at the \ion{Fe}{1} 6173\AA{}
absorption line with bandwidth $\Delta\lambda$ = 0.076\AA{},
spatial size scale of {0.5\arcsec} per pixel, and is obtained at
45-second cadence (with the exception of the vector magnetograms, obtained at
135-second cadence) \citep{Schou2012}.

Five continuous hours of \textit{SDO} observations centered on
SOL2011-02-15T01:56 were analyzed for this study. A C-flare occurred
before the X-flare between 00:30 and 00:45 UT, and two events occurred
after the X-flare, between 03:00 and 03:15, and between 04:25 and 04:45. (For
clarification, the main target of study (SOL2011-02-15T01:56) will be
explicitly referred to as the ``X-flare''.)

\begin{figure*}[htb!]\centering
    \includegraphics[width=1.0\textwidth]{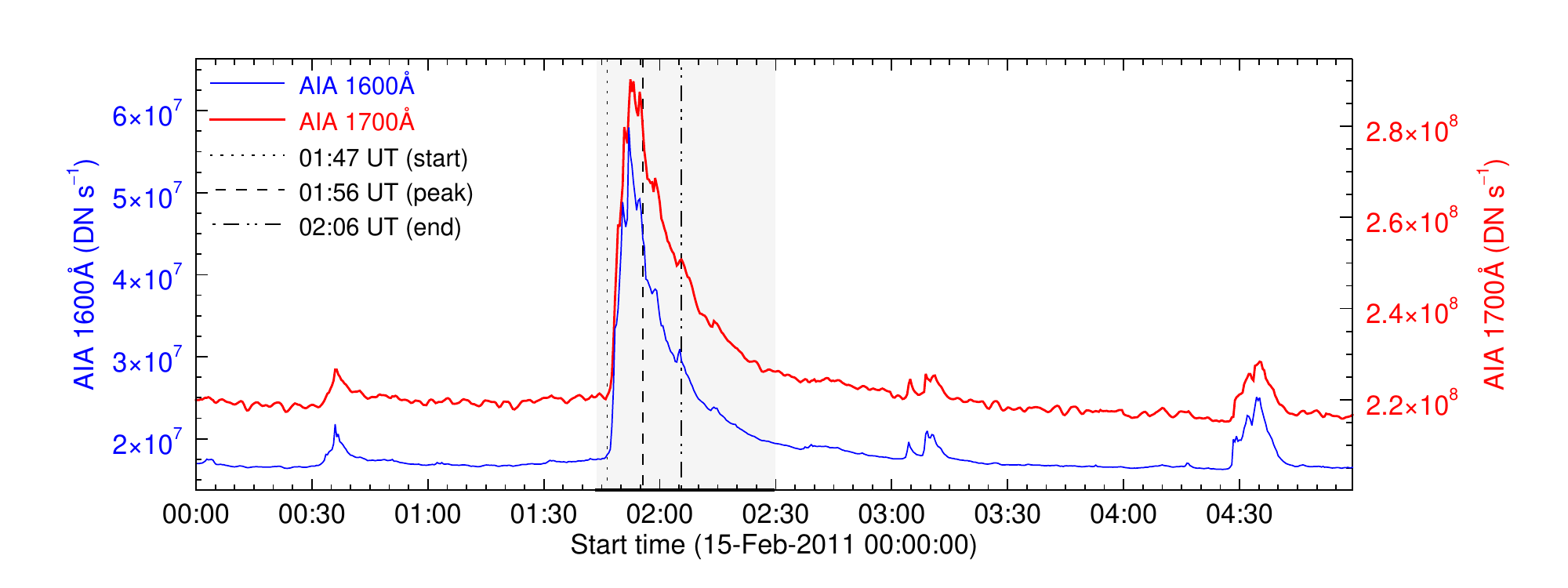}\\
    \includegraphics[width=1.0\textwidth]{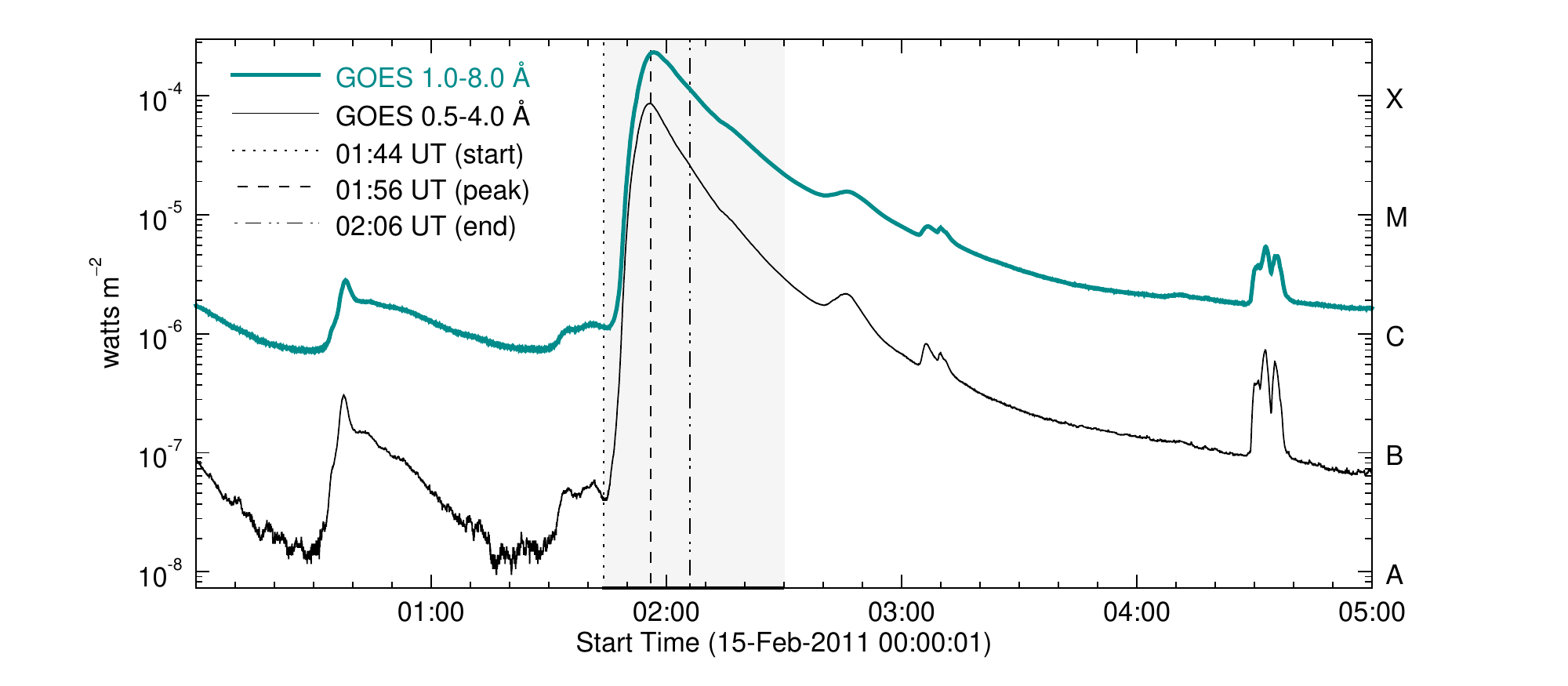}
    \caption{%
        Top: Light curves of the UV continuum emission from
        AIA 1600\AA{} (blue curve) and AIA 1700\AA{} (red curve),
        integrated over the
        {300\arcsec$\times$198\arcsec} subset of data centered on
        NOAA AR 11158.
        Bottom: Light curves observed by the \textit{GOES-15} satellite channels
        0.5-4\AA{} (black curve) and
        1-8\AA{} (green curve).
        The vertical lines mark the flare start, peak, and end times as
        determined by the 1-8\AA{} channel on \textit{GOES-15}.
        The shaded regions indicate the time period during the flare
        between 1:47 and 2:30. The unshaded regions on either side indicate the
        pre- and post-flare time periods.
        \label{lc}}
\end{figure*}

Figure~\ref{lc} shows the light curves
from 00:00 UT to 04:59 UT on 2011 February 15.
The shaded region indicates the
time period during the flare, between 1:47 and 2:30, which covers the flare
peak and the full duration of the particle injection event. The unshaded time
period from 00:00 to 01:47 illustrates the pre-flare emission, and the unshaded
time period from 02:30 to 04:59 illustrates the post-flare emission. Results
from these three phases are compared to analyze the flare patterns before,
during, and after the flare.

The emission from both AIA channels peaked just before 01:53 UT
(AIA 1600\AA{} peaked at 01:52:41.12, and AIA 1700\AA{} peaked at 01:52:55.71),
$\sim$3 minutes before the \textit{GOES} emission peaked at 01:56 UT.
Both AIA UV light curves show a slight increase near the \textit{GOES} end time
at 02:06 UT, followed by several additional increases during the decay
phase of the X-flare.
The standard data reduction routine \verb|aia_prep.pro| from SolarSoft
\citep{Freeland1998} was applied to the data from both instruments to co-align
the full disk images over all channels, de-rotate them, and set a common plate
scale of {0.6\arcsec} per pixel.
A {300\arcsec$\times$198\arcsec} subset of data centered on AR 11158 was extracted
from the full-disk set of images.
To correct
for solar rotation, the images were co-aligned
to a common reference image by cross correlation, providing sub-pixel accuracies
\citep{McAteer2003,McAteer2004}.
While several images were missing, the gaps were accounted for by
averaging the two adjacent images.

Both AIA channels saturated ($\geq\!15000$ counts) in the center of the AR
during the flare peak, resulting in pixel bleeding in the north and
south directions during the peak of the X-class flare. A few pixels also
saturated during smaller, surrounding events.
These pixels were all contained within the
{300\arcsec$\times$198\arcsec} subset of data throughout the
duration of the time series.
Where saturation occurs in the power maps,
the pixel value is set to zero.

\begin{figure}[htb!]\centering
    \includegraphics[width=0.5\textwidth]{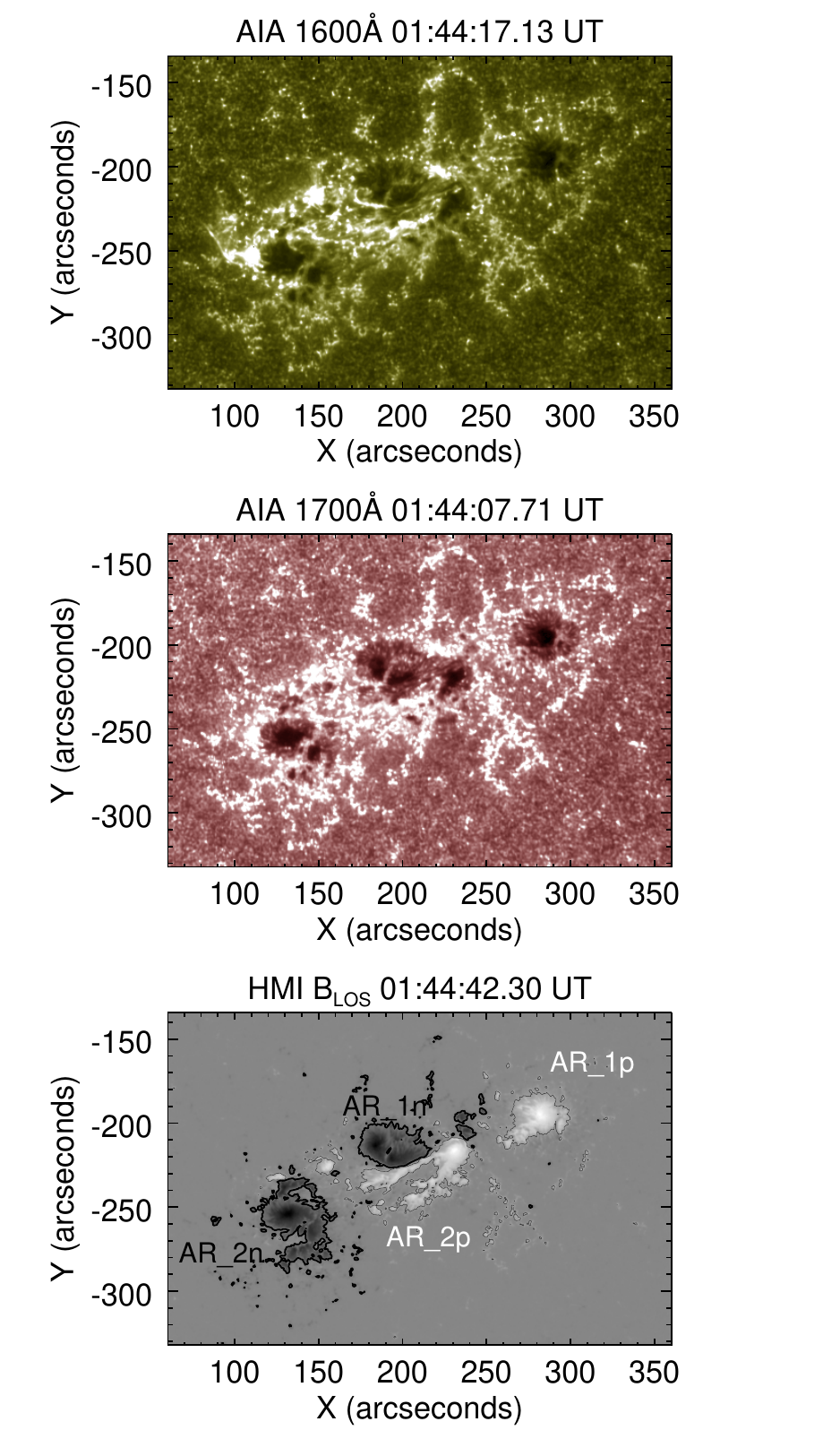}
    \caption{
        AR 11158 in AIA 1600\AA{} (top), AIA 1700\AA{} (middle), and
        HMI LOS magnetogram (bottom) on 2011 February 15.
        The thin and thick contours outline
        positive (+300 Gauss) and negative (-300 Gauss)
        polarities, respectively.
        The two sunspots in the northern pair are labeled AR\_1p (leading sunspot)
        and AR\_1n (trailing sunspot).
        The two sunspots in the southern pair are labeled AR\_2p (leading sunspot)
        and AR\_2n (trailing sunspot).
        \label{images}}
\end{figure}

Pre-flare images of AR 11158 are shown in Figure~\ref{images}. The HMI
magnetograms show the magnetic configuration of the quadrupole created by two
sunspot pairs. The northern pair will be designated as AR\_1, where the leading
positive spot is AR\_1p and the trailing negative spot is AR\_1n. Similarly,
the southern pair will be designated as AR\_2, where the leading positive spot
is AR\_2p and the trailing negative spot is AR\_2n.
Of the four events prior to and following the X-class flare, all but
one occurred in AR\_2n.

\section{Analysis}\label{analysis}

To study the spacial and temporal variations in 3-minute oscillatory power,
a technique similar to that presented by
\cite{Jackiewicz2013} and applied by \cite{Monsue2016} is employed.
The method is essentially a windowed Fourier transform (WFT).
Start with a data cube $f(x,y,t)$
consisting of $N$ intensity images separated by timestep $dt$
(= instrumental cadence), and spans a total duration = $N*dt$.
A subset of window length $T$ is extracted,
covering the time between $t_{i}$ and $t_{i}+T$,
A Fast Fourier Transform (FFT) is applied to every pixel $(x,y)$ in this subset
in the temporal direction.
The 3-minute power at that pixel is approximated from the Fourier power spectrum
by averaging the power over a frequency bin of width $\Delta{\nu}$
centered on the frequency of interest.
The process is repeated for
all $i = 0:(N*dt)-T$,
where each $i$ increment shifts the time window by one timestep $dt$.
The result is an array of power maps $P(x,y,t)$.

For this study,
the window length $T$ was set to 64 images ($\sim$25.6 minutes), and
the frequency bin $\Delta\nu$ was set to 1 mHz centered on $\nu \sim 5.6$ mHz.
Since power maps were computed starting at every possible value of $i$,
there is some overlap in the time periods from which adjacent maps were obtained.
This window length is necessary to achieve a frequency resolution $\delta\nu$
that allows at least two frequencies to fall within the
frequency bin $\Delta\nu$ in the power spectrum.
Averaging over more than one frequency point helps reduce the noise in the
resulting power maps, at the expense of resolution of the flare.
Two frequency points were obtained within $\Delta\nu$:
at 5.21 mHz (192.00 seconds) and 5.86 mHz (170.67 seconds).
The average power in the 3-minute mode was computed from these two points.

Each Fourier transform was applied without detrending the data since
the frequency of interest was well outside the global flare signal.
This was confirmed by applying a Fourier filter with a cutoff period
longer than 400 seconds.
The power spectra for the periods of interest was not altered.
If a saturated pixel was encountered during the FFT computation,
the value of that pixel was set to zero in the power map for that window.


\section{Results and Discussion}\label{results}

\subsection{Spatial distribution of 3-minute power}

Figures~\ref{before}, \ref{during}, and \ref{after} show intensity
images and corresponding spatial distribution of the 3-minute power in AIA
1600\AA{} and AIA 1700\AA{} before, during, and after the flare, respectively.

Throughout the 5-hour time series, enhanced 3-minute power
(spatially, pixels with higher 3-minute power than others)
is located over the interior or boundaries of the
umbral/penumbral regions of AR 11158.
Very little enhancement is observed in the quiet chromosphere
(examination of the isolated power maps outside the AR confirmed that this
observation was not the result of a visual scaling effect).
Suppression of the
3-minute power appears in areas directly over the center of the umbra, with a
ring-shaped pattern of enhanced power around the outer edges of the
umbra, consistent with results obtained by \cite{Reznikova2012} using data from
the 1600\AA{} and 1700\AA{} channels on \textit{SDO}/AIA.
The size of enhanced regions ranges between a few (4-6) pixels to
$\gtrsim 10$ pixels,
all smaller than the umbrae of the sunspots and consistent with the
size of flare footpoints observed in the chromosphere.
Locations of enhanced power in the maps often
appear to be loosely correlated
with locations of high intensity in the corresponding intensity images.
However, enhanced power did not appear in every location of high intensity.
This is most clearly visible in AIA 1600\AA{}.
The regions of enhanced 3-minute power were usually located along the
boundaries of magnetic field strength at $\pm$300 Gauss, approximately over the
outer penumbral boundary.
The small areas of enhanced power emerge and fade at a later time from the same
location, though the temporal resolution obtained does not allow the
determination of precise times of either.
One of the most prominent locations of enhanced power occurs before the flare
at the bottom of the leading sunspot in the northern pair.

\subsubsection{Pre-flare}\label{ssec:before}

\begin{figure*}[p]\centering
    \includegraphics[width=0.9\textwidth]{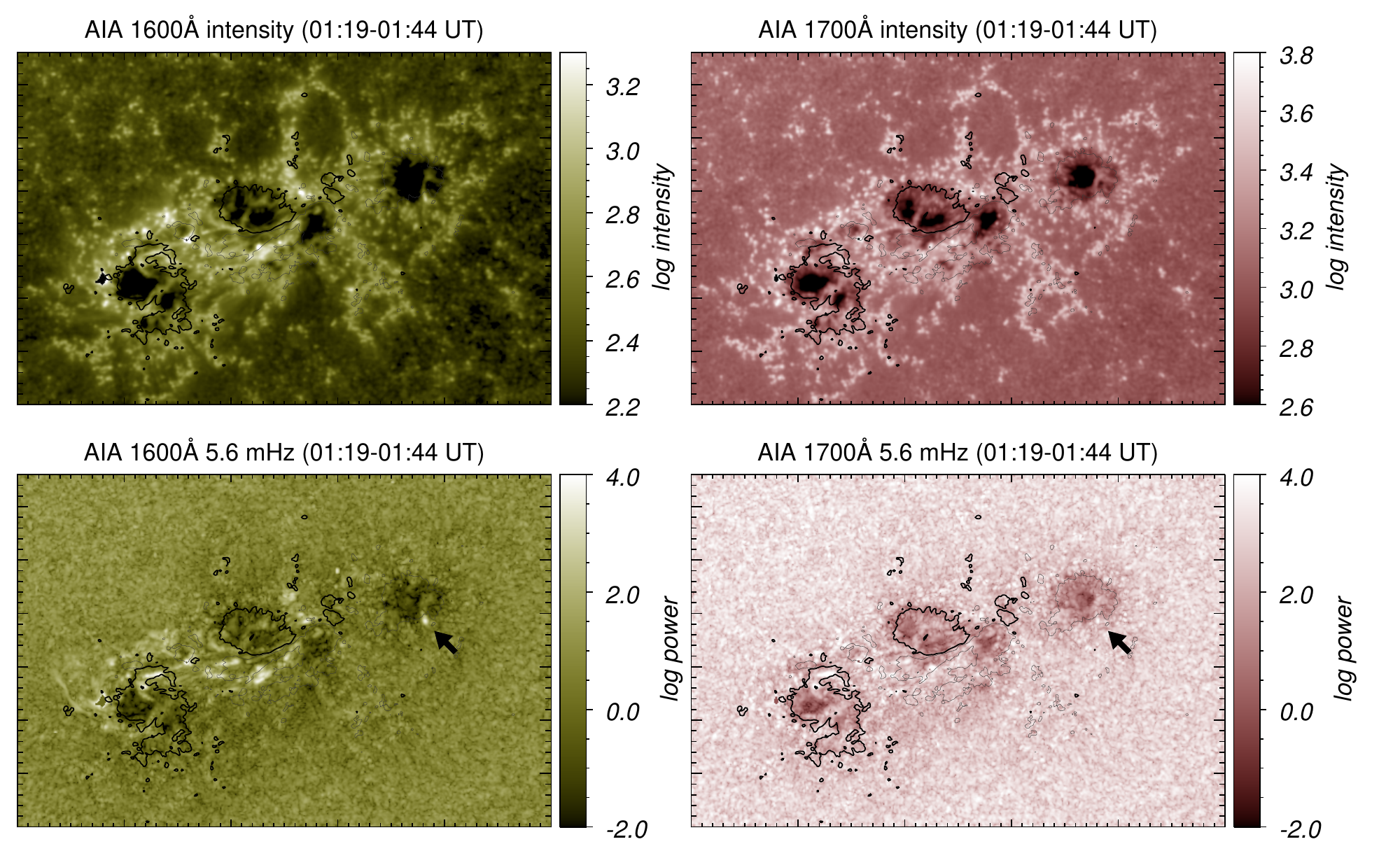}
    \caption{%
        Intensity and spatial distribution of 3-minute power immediately prior
        to the X-class flare between 01:19 and 01:44 UT on 2011 February 15,
        in log scale.
        Locations whose window included saturated pixels were set to zero.
        Contours indicate the approximate position of
        HMI B$_{LOS}$ at $\pm$300 Gauss.
        Thin and thick contours represent
        positive and negative polarities, respectively.
        The dimensions of each image are the same as labeled on the axes in
        Figure~\ref{images}.
        The arrow indicates the enhanced area discussed in the text.
        \label{before}}
\end{figure*}

Figure~\ref{before} shows intensity images and corresponding spatial
distribution of the 3-minute power in AIA 1600\AA{} and AIA 1700\AA{} obtained
from observations immediately prior to the flare (01:19-01:44).
The contours represent the line-of-sight magnetic field (HMI $B_{LOS}$) at
$\pm$300 Gauss. The contours indicate the approximate location of the boundary
of the outer penumbra.
Both the contour data and the
intensity maps were obtained by averaging over the window from which the power
map was computed. The alignment procedures may have resulted in a slight offset
between the channels, so the position of the contours is approximate.

Locations of enhanced power are, for the most part, located along
umbra/penumbra boundaries.
In Figure~\ref{before}, there is a bright spot with high power along the
southern penumbra boundary of AR\_1p, with an area of approximately
{6\arcsec$\times$6\arcsec}.
This is indicated by the black arrow in the bottom row of
Figure~\ref{before}.
The time series of AR intensity leading up to the X-flare reveals
intermittent appearances of high intensity at this location.
    This enhancement in power appeared in
    power maps obtained from observations times
    as early as 40 minutes prior to flare start.
Due to the length of $T$,
    it is difficult to resolve the temporal evolution of this region.
    However, since it is present in pre-flare power maps, it
    is possible that the power enhancement is caused by
    by non-thermal particles whose emission at this time was not yet strong
    enough for \textit{GOES} flare start time.

    A
C-class flare
occurred an hour before the X-flare, between 00:30 and 00:45 UT.
The emission from the C-flare originated from AR\_2n,
on the opposite side of AR 11158
from the bright region discussed in the previous paragraph.
    Though Figure~\ref{before} does not cover this time period,
    it remains possible that post-flare effects from the C-flare may be present
    in the pre-flare data for the X-flare.

\subsubsection{During the flare}\label{ssec:during}

\begin{figure*}[p]\centering
    \includegraphics[width=0.9\textwidth]{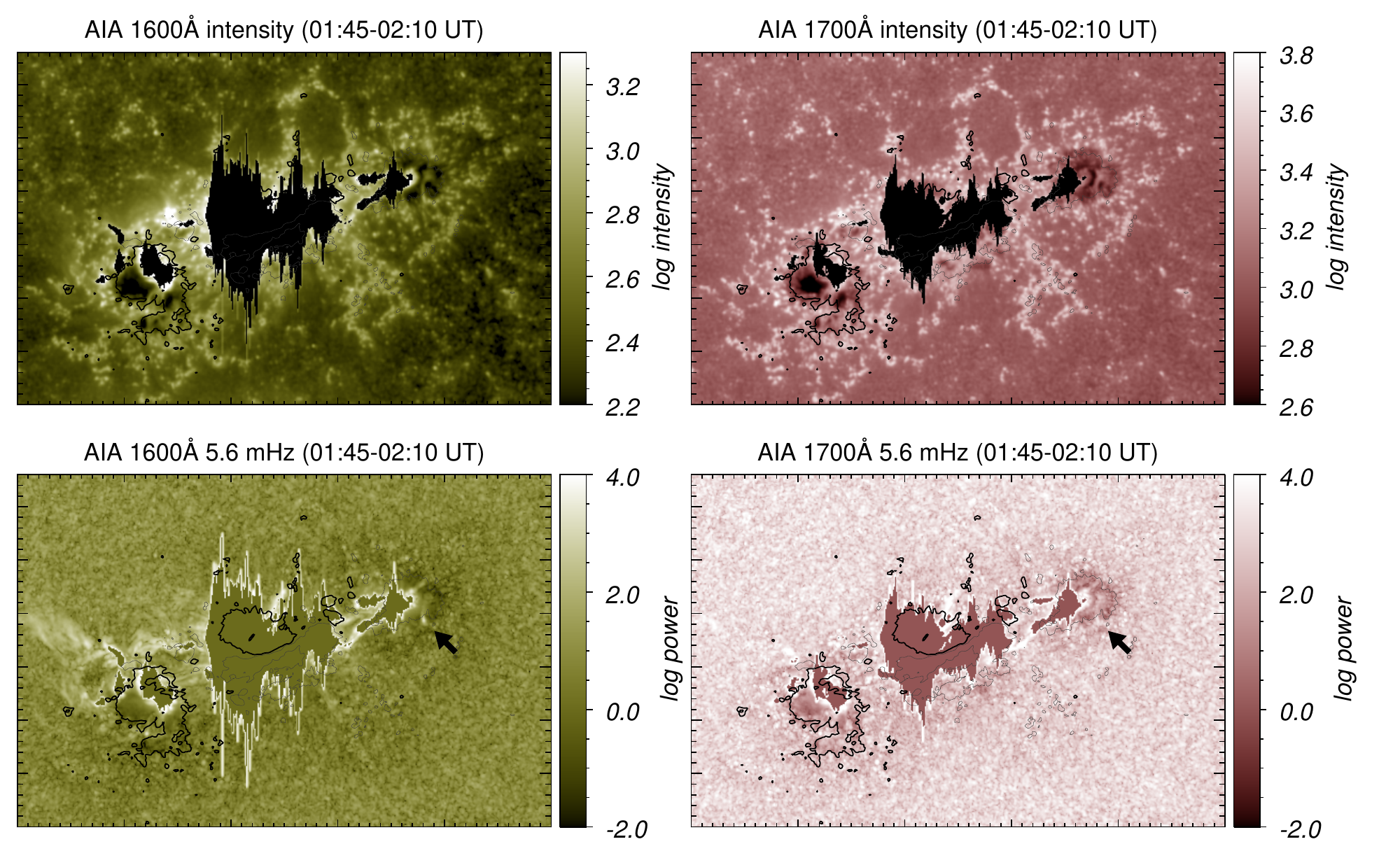}
    \caption{%
        Same as Figure~\ref{before}, during the X-class flare between
        01:45 and 02:10 UT on 2011 February 15.
        \label{during}}
\end{figure*}

Figure~\ref{during} shows intensity images and corresponding power maps
obtained from observations starting at the impulsive phase,
between 01:45 and 02:10 UT.
Saturated pixels were excluded from the maps.

The region of enhanced power along the southern penumbral boundary
of AR\_1p discussed in \S\ref{ssec:before}
is seen in maps covering
the precursor and impulsive phases of the X-flare,
but does not appear in power maps obtained from the gradual phase.
This may suggest this region was responding to energy input via non-thermal
particles, since magnetic reconnection and particle acceleration take place
during the impulsive phase.

The concentration of power enhancement in small areas located in the vicinity
of sunspot umbrae suggests that these small areas of the chromosphere are
responding directly to the injection of energy by a beam of non-thermal
particles. Response of deeper layers in the solar atmosphere were
detected via impact points localized in space in the photosphere along the
outer penumbral boundary during the impulsive phase of the same flare. These
sunquakes were suggested to have been triggered at the footpoints of the flux
tube when it was released at the beginning of the impulsive phase
\citep{Kosovichev2011, Zharkov2011}.

If the location of power enhancement reveals source locations of energy
injection via non-thermal particle beams,
then it remains possible that changes in spatial locations of these sources
implies a change in the site where magnetic reconnection takes place in the
corona.
In fact, \cite{Kuroda2015} reported two distinct magnetic reconnection
sites during the impulsive phase of the 2011 February 15 flare.
Due to the saturation of pixels in AIA UV data at this time,
it cannot be confirmed here whether the impulsive
phase shows enhancement of oscillatory power at the same locations.

Since a large number of pixels saturated during the flare, there is little
spatial information that can be extracted from power maps obtained during the
flare, particularly from the AR center and parts of the outside sunspots, where
emission reaches further into as the flare develops.
While QPPs have been observed to persist for as many as 163 pulses during
an 8-hour flare with periods between 10 and 100 seconds \citep{Dennis2017},
QPPs in flare emission often do not persist longer than three or four periods,
so extracting these oscillations would be difficult even without saturation
with the current methods (see review from \cite{Inglis2009}).

\subsubsection{Post-flare}\label{ssec:after}

\begin{figure*}[p]\centering
    \includegraphics[width=0.9\textwidth]{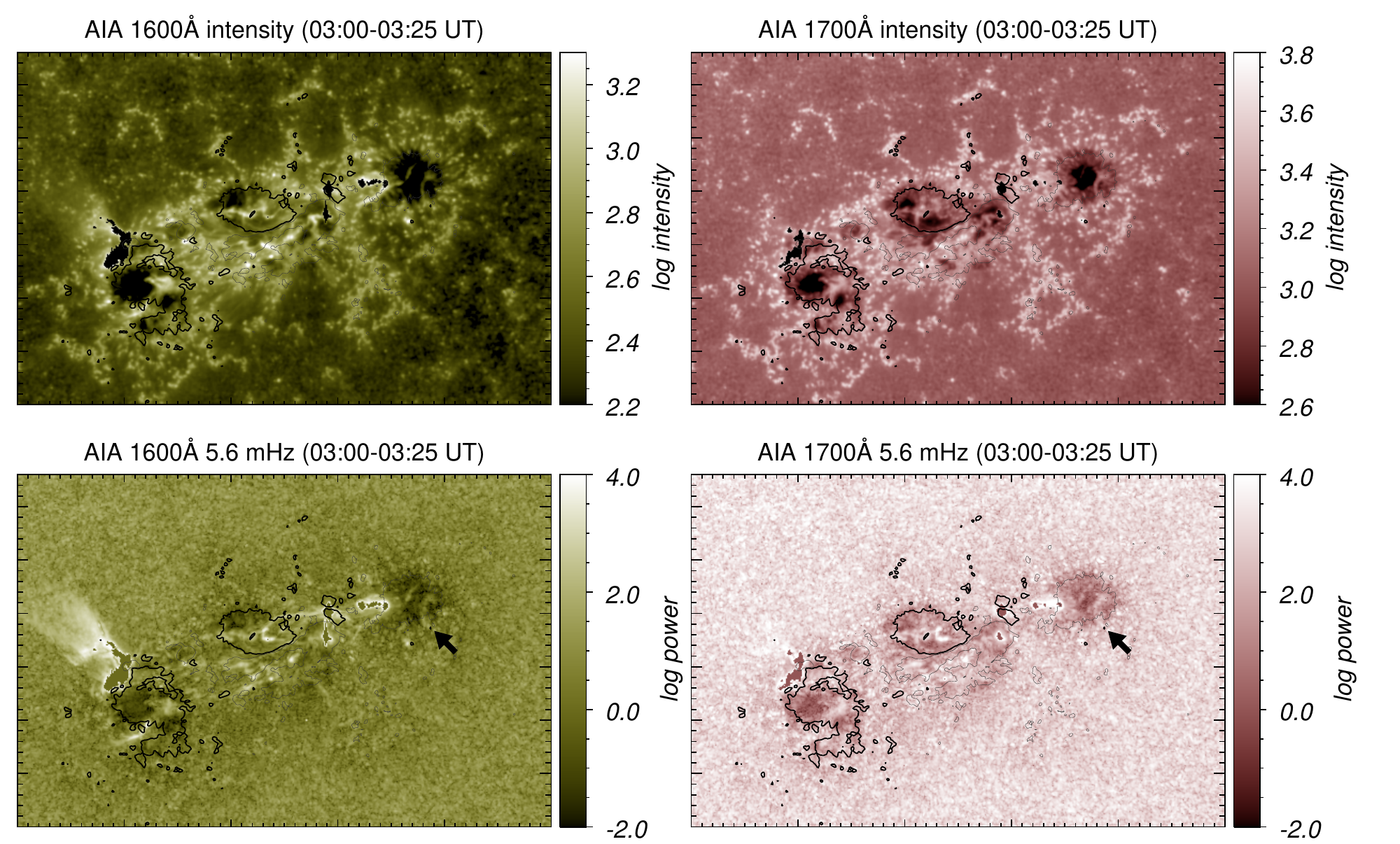}
    \caption{%
        Same as Figure~\ref{before}, after the X-class flare between
        03:00 and 03:25 UT on 2011 February 15.
        \label{after}}
\end{figure*}

Figure~\ref{after} shows intensity images and corresponding power maps
obtained from observations between 03:00 and 03:25.
This time window includes the
first of two small events that occurred after the X-class flare.
During this post-flare time period,
oscillatory power enhancements are present at several small regions in the
center, right around AR\_2p.

\subsection{Temporal evolution of 3-minute power}

\begin{figure*}[p]\centering
    \includegraphics[width=\textwidth]{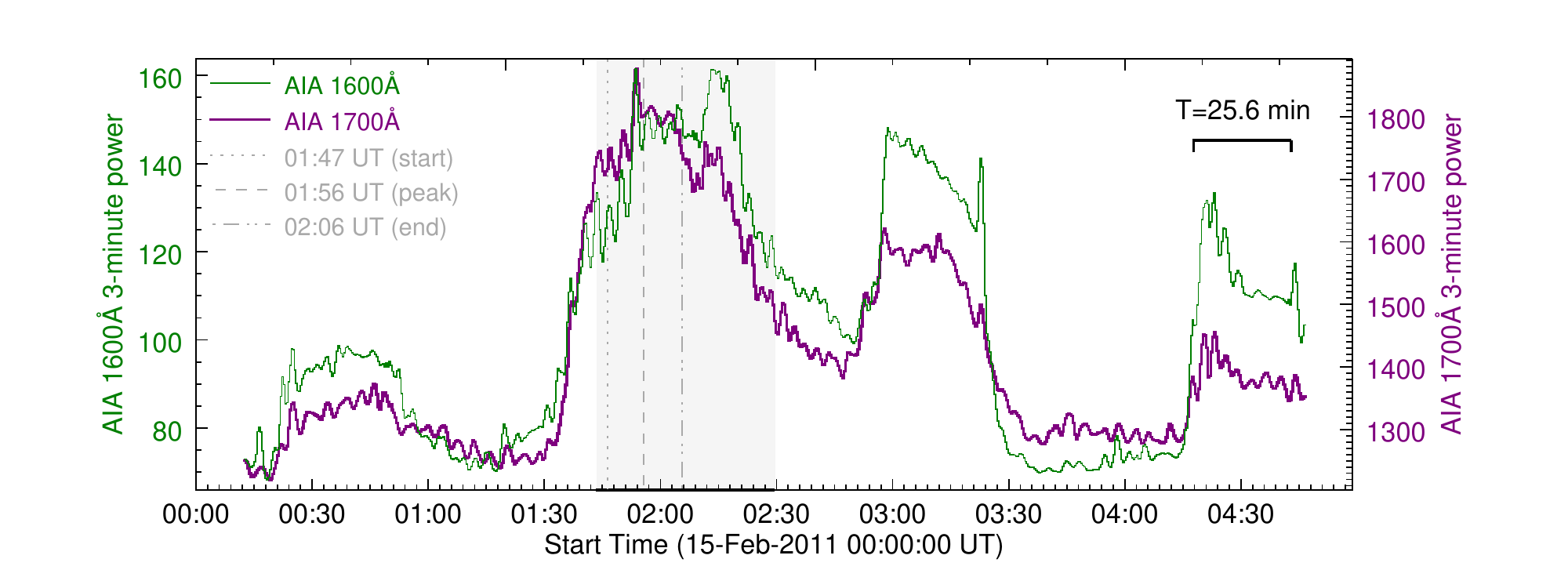}
    \caption{%
        Temporal evolution of the 3-minute power $P(t)$
        per unsaturated pixel
        in AIA 1600\AA{} (green curve) and AIA 1700\AA{} (purple curve),
        obtained by summing over each {300\arcsec$\times$198\arcsec}
        power map.
        Each point in time is plotted at the center of the window over which the
        Fourier transform was applied to obtain the power map.
        The vertical dashed lines mark the \textit{GOES} start, peak, and end
        times of the flare at 01:44, 01:56, and 02:06 UT, respectively.
        The shaded region covers the time period during the flare
        (see Figure~\ref{lc} caption).
        \label{pt}}
\end{figure*}

The temporal evolution of 3-minute power was estimated by reducing the 3D cube
of power maps to a 1D array through time by summing each map \boldmath $
P_{i}(x,y,\nu) $ over $x$ and $y$ and taking the total to be the 3-minute power
from the $500\times330$ pixel subset of data centered on AR 11158.
Integrating flux
over the AR before applying the Fourier transform has the possible effect of
reducing or canceling signal from pixels whose intensity variations are out of
phase, or enhancing signal from pixels whose intensity variations are in phase.
Figure~\ref{pt} shows the total 3-minute power with time, obtained by shifting
the array by $T/2$ so that each point is plotted at the center of the time
window from which its value was obtained.
The axis labelled `$T=25.6$ min' is presented to illustrate the
duration of the window $T$.
Since each map was obtained from flux signal spanning a time window of length
$T$ = 25.6 minutes, each point in the 1D array represents power
from $t_{i}$ to $t_{i}+25.6$.

Since the numerical values were calculated from non-physical numbers
(DN s$^{-1}$), only the relative changes
and overall pattern for each channel are considered for interpretation.

After the peak in power following the peak in emission
(just after \textit{GOES} peak time),
there is a second peak (local maximum) in the
3-minute power between $\sim$2:10 and $\sim$2:17 UT.
The light curves show several peaks during the decay phase.

An increase in total emission from an AR during this phase in thermal UV
wavelengths could indicate a different source of energy input. Since
acceleration of non-thermal particles is no longer taking place, this could be
conversion to thermal energy from another form. There may also be a change in
formation height of the emission observed by each channel as flare evolves.

As observed previously by \cite{Milligan2017}, the 3-minute power increases
during the X-flare, as well as during the smaller events before and after, as
shown in Figure~\ref{pt}.
The persistence of the 3-minute power toward
the end of the gradual phase in AIA 1700\AA{} is consistent with the results of
the wavelet analysis carried out by \cite{Milligan2017}.

The amplitude of the small-scale variations in 3-minute power
is higher for AIA 1700\AA{} almost everywhere with
the exception of the main phase of the X-class flare.
The power from AIA 1700 is higher than AIA 1600 by
1000 counts at all points throughout the time series.
Compared to 1700\AA{},
the 1600\AA{} 3-minute power appears to increase more
(relative to its own minimum) and at a faster rate.
The standard deviation for $P(t)$ from integrated flux
for 1600\AA{} is $4.9 \times 10^{4}$, and
for 1700\AA{} is $2.7 \times 10^{4}$.

If the emission from AIA 1600\AA{} originates from a higher location in the
atmosphere than the 1700\AA{} emission, a possible explanation for the higher,
sharper increase is that the energy from the non-thermal particle beam
dissipates as it travels through deeper layers of the chromosphere. Although
the emission from AIA 1700\AA{} is generally thought to originate in lower
formation heights than emission from 1600\AA{}, the latter spans a broader
temperature range, and contains emission from the \ion{C}{4} line.
Determination of the AIA 1600\AA{} formation height is more complicated during
flares because the \ion{C}{4} is more likely to be contributing to the signal,
and both channels may be sampling at deeper layers than they are thought to
during non-flaring times.

\cite{Simoes2019} investigated the spectral contribution of these two channels
during flares, using high-resolution spectra from Skylab to provide reference
spectra from plage observations during quiescent times.
They found that flare excess emission is chromospheric in origin, and is
dominated by contribution from spectral lines, while the quiet chromosphere is
dominated by continuum emission.

Due to the relatively long window contributing
to the power at each point in time for Figure~\ref{pt},
the time-resolution is rather poor,
making it
difficult to determine the precise flare phase during which power changed.

\subsection{Time-frequency power plots}

\begin{figure*}[p]\centering
    \includegraphics[width=0.9\textwidth]{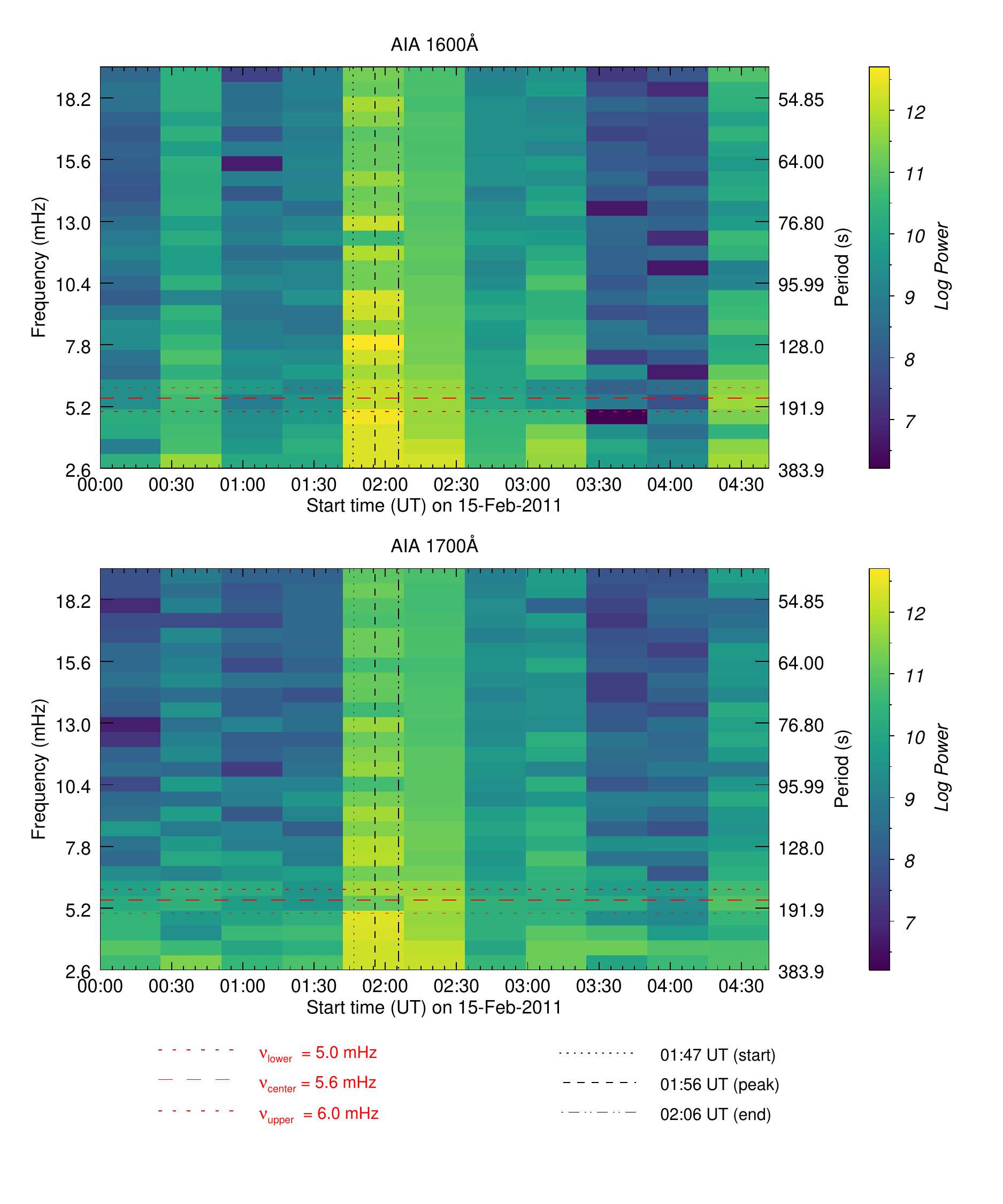}
    \caption{
        Time-frequency power plots from AIA 1600\AA{} (top panel) and AIA
        1700\AA{} (bottom panel), obtained by applying a Fourier transform to
        integrated emission over the {300\arcsec$\times$198\arcsec}
        subset of data centered on NOAA AR 11158 in discrete time increments
        of 64 frames ($\sim$25.6 minutes) each with no overlap (i.e.
        for start time $t_{0}$, then $t_{1} = t_{0}+T$, etc.). The dashed
        horizontal line marks the central frequency $\nu_{c}$ at $\sim$5.6 mHz,
        corresponding to a period of 3 minutes. The dotted horizontal lines on
        either side of $\nu_{c}$ mark the edges of the frequency bandpass
        $\Delta\nu$ = 1 mHz. The vertical lines mark the flare start, peak, and
        end times as determined by \textit{GOES}. The power is scaled
        logarithmically and over the same range in both channels. Note
        that unlike Figure~\ref{pt}, the emission was summed over all pixels in
        the $x$ and $y$ dimensions prior to applying the Fourier transform,
        resulting in power values much higher than the range displayed for
        Figure~\ref{pt}, as power scales as the square of the input signal.
        \label{wa}}
\end{figure*}

In addition, the resulting power is significantly lower when the
Fourier transform is applied to each pixel before summing over all $x$ and $y$
in the subset, as power scales as the square of the input signal. The following
section contains results computed by first integrating the flux and the power
is significantly higher, as seen in Figure~\ref{wa}.

The technique described in \S\ref{analysis} was applied to the integrated flux
from AR 11158
over the {300\arcsec$\times$198\arcsec} subset used to compute the power
maps in Figures~\ref{before}, \ref{during} and \ref{after} and generate the
plot of 3-minute power with time, shown in Figure~\ref{pt}. Instead of
advancing in intervals equal to the instrumental cadence, the technique was
applied
at discrete intervals of $T=64$ images ($\sim$25.6 minutes)
with no overlap (i.e. for start time $t_{0}$, then $t_{1} = t_{0}+T$, etc.)
This technique is similar to one employed by \cite{Monsue2016}
during two flares, here applied to an extended observation time
of the flare studied by \cite{Milligan2017}.
It provides a way to
compare the spectral power at a range of frequencies.
The results are similar
to those obtained with wavelet analysis, with power at a range of frequencies,
each shown as a function of time, but with lower frequency and time resolution.
The results are are shown in Figure~\ref{wa} for frequencies between 2.5 and
20.0 mHz (400 and 50 seconds, respectively).
The central frequency $\nu_{c} =
5.56$ mHz and the frequency bandpass $\Delta\nu=1$ mHz between 5 and 6 mHz are
marked by the horizontal dashed lines.

At all points in time when power enhancement occurs for any frequency,
there appears to be a correlation with flux increase, as illustrated in the
light curves in Figure~\ref{lc}, namely during the following time ranges:
00:30-00:45, 01:45-02:30, 03:00-03:15, and 04:20-04:45 UT. The power at all
frequencies is enhanced in these plots during the X-flare compared to
their non-flaring power before and after.
This implies that the chromospheric plasma oscillates at a range of frequencies
in response to energy injection.
During the small events before and after the flare, the power at lower
frequencies is enhanced, but the power at higher frequencies is
suppressed relative to the same frequencies for adjacent windows.


\section{Conclusions}\label{conclusions}

We have presented the spatial distribution of the 3-minute oscillations
associated with the X-class flare that occurred in AR 11158 on 2011 February 15.
Key points are as follows:
\begin{enumerate}
    \item Concentrated regions of 3-minute power
        indicate that the chromospheric plasma does not oscillate as one body.
    \item The locations of increased 3-minute power during the flare
        are co-spatial with those of HXR flare emission, concentrated in areas
        $\sim$10 pixels ($\sim4\arcsec$) across. This suggests that these small
        areas are manifestations of the chromosphere responding to injection of
        energy by nonthermal particles, and supports the theory that the
        chromosphere oscillates at the acoustic cutoff frequency in response to
        a disturbance.
    \item Variation in enhancement location throughout flare phases
        indicates a possible change in reconnection site location from which
        particles are accelerated.
\end{enumerate}

The temporal behavior of oscillations during the main flare remains
inconclusive due to the necessary balance between temporal and frequency
resolution.
Techniques to improve temporal resolution, such as the standard
wavelet analysis presented by \cite{Torrence1998},
will be used in future analysis to
analyze additional flares (including pre- and post-flare phases) to allow the
study of chromospheric behavior on timescales comparable to those over which
flare dynamics are known to occur.
Future work will include multiple flares smaller than X-class to
reduce the number of pixels contaminated by saturation and
obtain more conclusive results from the flare core region.
Since the pre-flare data shown in this work includes a C-flare, it may be
worthwhile to obtain more data at earlier times.
Hinode/SOT observed the 2011 February 15 X-flare without saturation
\citep{Kerr2014};
these data will be considered for future analysis of this flare.


\acknowledgments{}
The authors gratefully acknowledge
the anonymous referee for comments and suggestions improving this work.
We also thank Sean Sellers, Sarah Kovac, and Jason Jackiewicz
for help with the text.
This work was supported by funding provided via NSF \# 1255024.

\bibliography{reffile}

\end{document}